\documentclass{article}
\usepackage[utf8]{inputenc}
\usepackage{color}
\usepackage{cite}
\usepackage{url}
\usepackage{graphicx}
\usepackage{authblk}

\title{Note on Attacking Object Detectors with \\ Adversarial Stickers}
\author[2]{Kevin Eykholt}
\author[3]{Ivan Evtimov}
\author[3]{Earlence Fernandes}
\author[1]{Bo Li}
\author[1]{Dawn Song}
\author[3]{Tadayoshi Kohno}
\author[4]{Amir Rahmati}
\author[2]{Atul Prakash}
\author[5]{Florian Tramer}

\affil[1]{University of California, Berkeley}
\affil[2]{University of Michigan Ann Arbor}
\affil[3]{University of Washington}
\affil[4]{Stony Brook University}
\affil[5]{Stanford University}
\date{}

\begin{document}

\maketitle

\begin{abstract}

Deep learning has proven to be a powerful tool for computer vision and has seen widespread adoption for numerous tasks. However, deep learning algorithms are known to be vulnerable to \emph{adversarial examples}. These adversarial inputs are created such that, when provided to a deep learning algorithm, they are very likely to be mislabeled. This can be problematic when deep learning is used to assist in safety critical decisions. Recent research has shown that classifiers can be attacked by physical adversarial examples under various physical conditions.  Given the fact that state-of-the-art objection {\em detection} algorithms are harder to be fooled by the same set of adversarial examples, here we show that these detectors can also be attacked by physical adversarial examples. In this note, we briefly show both static and dynamic test results. We design an algorithm that produces physical adversarial inputs, which can fool the YOLO object detector and can also attack Faster-RCNN with relatively high success rate based on transferability. Furthermore, our algorithm can compress the size of the adversarial inputs to stickers that, when attached to the targeted object, result in the detector either mislabeling or not detecting the object a high percentage of the time. This note provides a small set of results. Our upcoming paper will contain a thorough evaluation on other object detectors, and will present the algorithm.

%Deep neural networks (DNNs) have enabled great progress in a variety of application areas, including image processing, text analysis, and speech recognition. DNNs are also being incorporated as an important component in many cyber-physical systems. For instance, the vision system of a self-driving car can take advantage of DNNs to better recognize pedestrians, vehicles, and road signs. However, recent research has shown that DNNs are vulnerable to adversarial examples: Adding carefully crafted adversarial perturbations to the inputs can mislead the target DNN into mislabeling them during run time. Such adversarial examples raise security and safety concerns when applying DNNs in the real world. For example, adversarially perturbed inputs could mislead the perceptual systems of an autonomous vehicle into misclassifying road signs, with potentially catastrophic consequences. 

%There have been several techniques proposed to generate adversarial examples and to defend against them. In this blog post we will briefly introduce state-of-the-art algorithms to generate digital adversarial examples, and discuss our algorithm to generate physical adversarial examples on real objects under varying environmental conditions. We will also provide an update on our efforts to generate physical adversarial examples for object detectors.

\end{abstract}

\section{Introduction}

Deep learning and deep neural networks(DNN) are widely used in computer vision, especially for safety critical tasks such as autonomous driving~\cite{lillicrap2015continuous}. However, research has shown that deep networks are also vulnerable to digital manipulations of their inputs, resulting in a class of attacks known as `adversarial inputs'~\cite{goodfellow2014explaining,papernot2016limitations,carlini2017towards,sabour2015adversarial,kos2017adversarial}. These digital manipulations cause deep networks to mislabel an input, despite no visible difference in the image. For example, a slight modification to some pixels in Stop sign image can cause a network to label it as a Speed Limit sign. 

%Recent research suggests that adversarial inputs are not a concern because current adversarial generation algorithms do not produce physical inputs that fool state-of-the art object detection algorithms~\cite{noneed}.

\begin{figure}[t]
  \centering
  \includegraphics[width=0.35\textwidth]{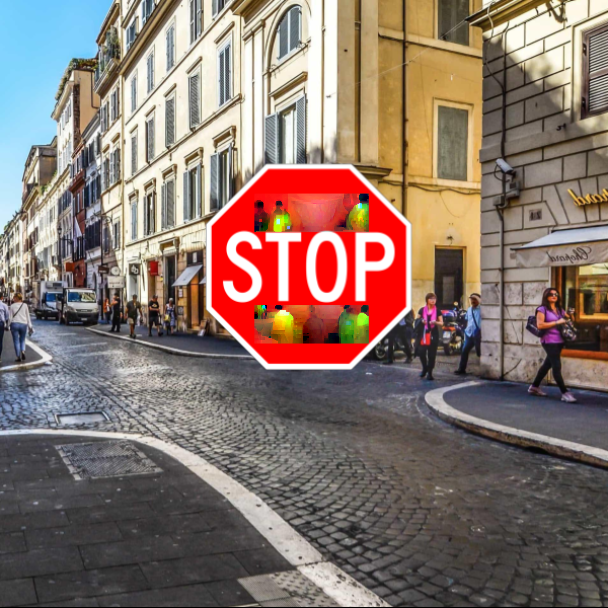}
  \caption{Output of our adversarial attack algorithm. The Stop sign in the image is printed such that is is the same size as a U.S. Stop sign.}
  \label{fig:noise}
\end{figure}

Our recent work introduced an algorithm that produced physical adversarial inputs, which fooled a DNN classifier trained for traffic sign recognition~\cite{evtimov2017robust}. Recently, we have also designed an algorithm that produces physical adversarial inputs, which can fool a state-of-the art object detection algorithm, YOLO (You Only Look Once). YOLO is a deep convolution neural network that performs real-time object detection~\cite{yolo}. Our algorithm can generate an entire object that is either mislabeled or not detected by YOLO. Also, the algorithm can generate small sticker perturbations that can be applied the object to similarly fool YOLO. See Figure \ref{fig:noise} for an example output of our algorithm generating an adversarial stop sign. We crop out the background of the generated image and print the Stop sign. For adversarial stickers, we cut out the perturbation areas and attach them a real Stop sign(e.g. in Figure \ref{fig:noise}, the sign areas would be the upper and lower colored rectangles on the sign). Compared with classifiers, detectors are more challenging to fool as they process the entire image and can use contextual information (e.g. the orientation and position of the target object in the scene) in their predictions.

% \section{Adversarial Object Generation}
% We will describe the algorithm to generate adversarial objects and stickers in a future update.

\section{Attack Demo}

\begin{figure}[h!]
  \centering
  \includegraphics[width=0.35\textwidth]{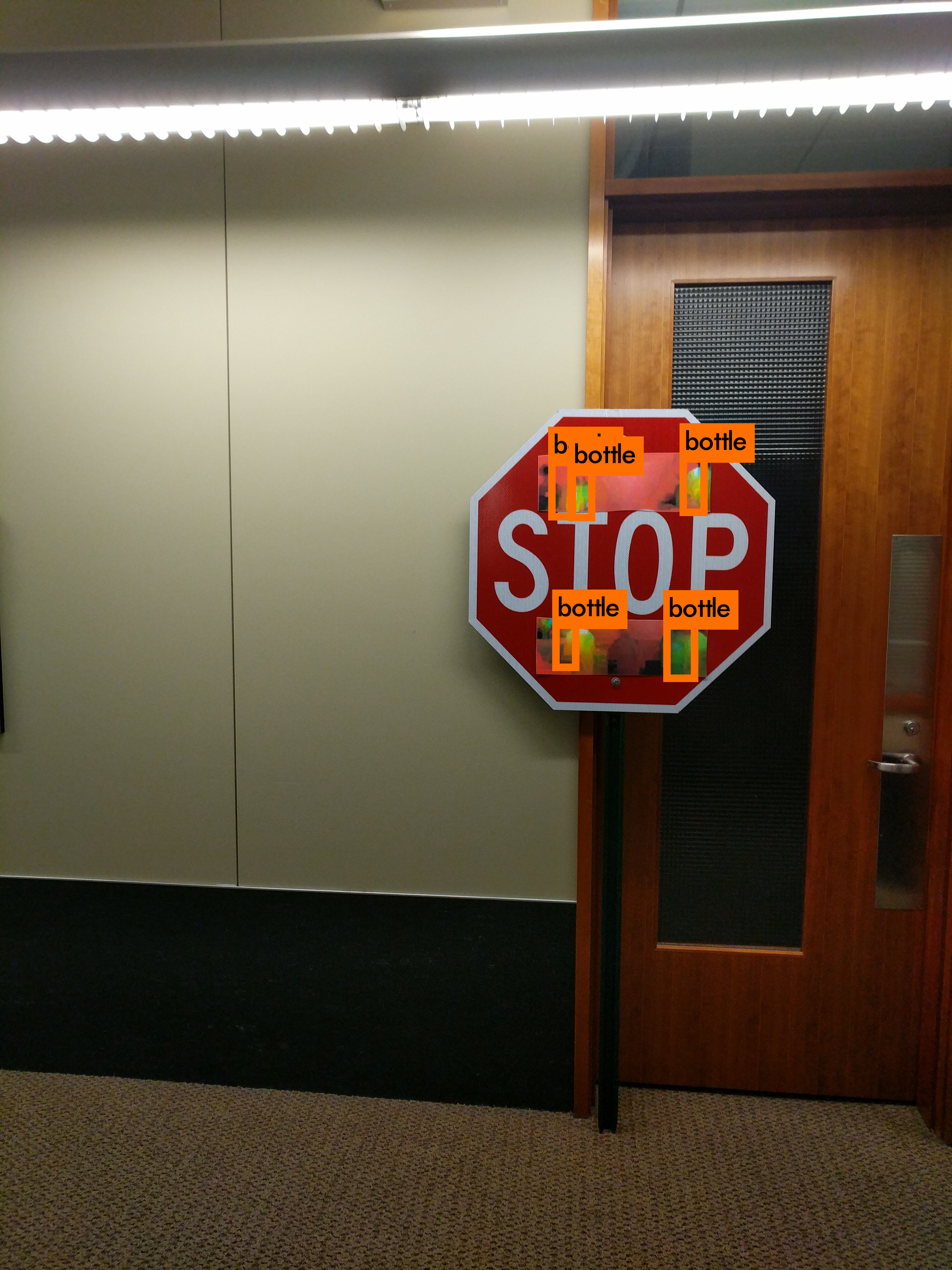}
  %\hfill
  \includegraphics[width=0.35\textwidth]{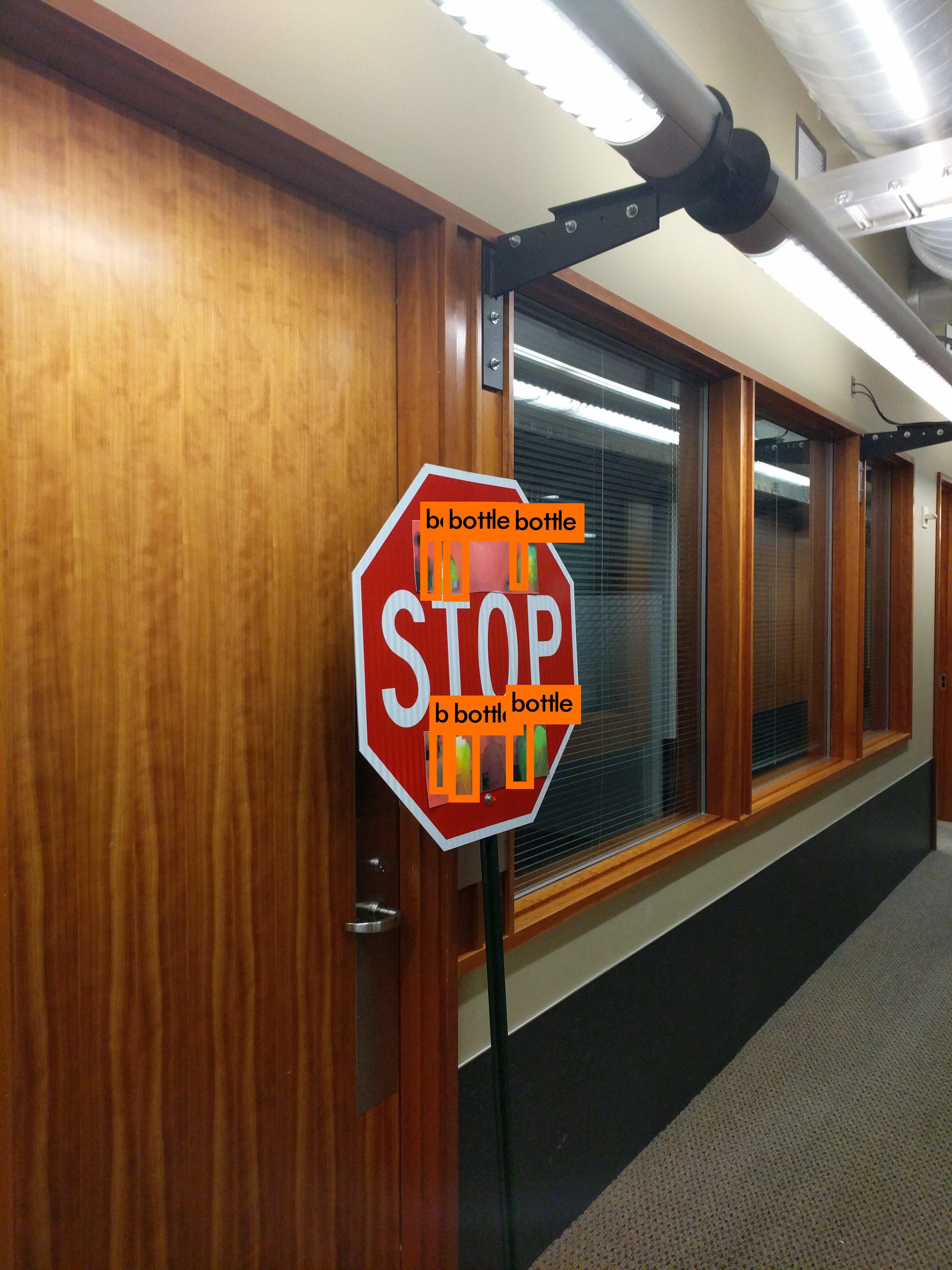}
  \caption{YOLO detection output for a real stop sign with adversarial stickers (phone camera pictures)}
  \label{fig:sticker-camera}
\end{figure}

\begin{figure}[h!]
  \centering
  \includegraphics[width=0.35\textwidth]{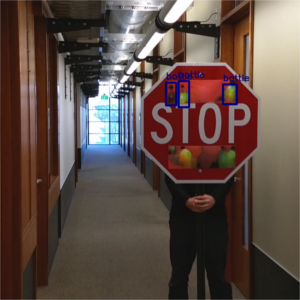}
  %\hfill
  \includegraphics[width=0.35\textwidth]{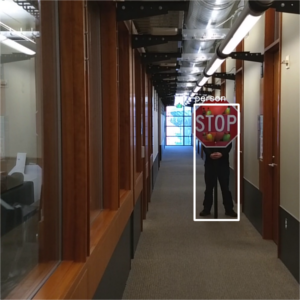}
  \caption{YOLO detection output for a real stop sign with adversarial stickers (still frames from video)}
  \label{fig:sticker}
\end{figure}

We evaluated our sticker attack against YOLOv2, which is trained to recognize 80 object classes including a class for Stop signs. For the experiment, we took several pictures of the object, a Stop sign, as well as a short video. These were provided to YOLO for detection. We used YOLO's default detection threshold of 25\% (i.e. an object is detected if YOLO is at least 25\% confident). Figure \ref{fig:sticker} and Figure \ref{fig:sticker-camera} show several images from our experiment as well as the detected objects in those images. For video results, please visit \textcolor{blue}{ \url{https://iotsecurity.eecs.umich.edu/#yolo}}. In the video, YOLO only detects the Stop sign when the camera is close to the sign. In an autonomous driving scenario, detecting a stop sign only a few feet from the sign is likely too late to react to the sign and safely brake.

In an effort to explore the transferability of our adversarial sticker generation algorithm, we provided the video to Faster-RCNN, another state-of-the-art object detector. Our adversarial stickers were able to fool Faster-RCNN in many of the video frames. Visit \textcolor{blue}{ \url{https://iotsecurity.eecs.umich.edu/#yolo}} to view the video results.  Given that this is a black-box attack on Faster-RCNN, it is expected that the attack may not be as successful on Faster-RCNN as it is on YOLO. In a future update, we plan to explore these details further.

\section{Algorithm Overview}

This algorithm is based off our earlier work on attacking classifiers~\cite{evtimov2017robust}. Fundamentally, we take an optimization approach to generating adversarial examples. However, our experimental experience indicates that generating robust physical adversarial examples for detectors requires simulating a larger set of varying physical conditions than what is needed to fool classifiers. This is likely because a detector takes much more contextual information into account while generating predictions. Key properties of the algorithm include the ability to specify sequences of physical condition simulations, and the ability to specify the translation invariance property. That is, a perturbation should be effective no matter where the target object is situated within the scene. As an object can move around freely in the scene depending on the viewer, perturbations not optimized for this property will likely break when the object moves. Our upcoming paper on this topic will contain more details on this algorithm.

\bibliographystyle{ieee}
\bibliography{reference}
\end{document}